\documentclass[letterpaper, 10 pt, conference]{ieeeconf}  

\IEEEoverridecommandlockouts                              

\usepackage{graphicx}
\usepackage{color}
\usepackage{colortbl}
\usepackage{multirow}
\usepackage{pifont}
\usepackage[]{footmisc}

\title{\LARGE \bf
Assessing  deep  learning methods  for  the  identification \\ of kidney  stones in endoscopic images*}

\author{Francisco Lopez$^{1}$, Andres Varela$^{1}$, Oscar Hinojosa$^{1}$, Mauricio Mendez$^{6}$, Dinh-Hoan Trinh$^{7}$,  \\ Jonathan ElBeze$^{4}$, Jacques Hubert$^{4,5}$, Vincent Estrade$^{3}$, Miguel Gonzalez$^{6}$,   Gilberto Ochoa$^{6}$, Christian Daul$^{2}$

\thanks{$^{1}$IEEE Student Member, INAOE, UAG (Mexico), UP (Colombia)}%
\thanks{$^{2}$CRAN (Universit\'e de Lorraine and CNRS), F-54000 Nancy}%
\thanks{$^{3}$CHU Pellegrin place Amémlie Raba Léon F-33000 Bordeaux}%
\thanks{$^{4}$CHU Nancy, Service d'urologie de Brabois, F-54511 Nancy}%
\thanks{$^{5}$IADI-UL-INSERM (U1254), F-54511 Vand{\oe}uvre-l\`es-Nancy}%
\thanks{$^{6}$Tecnologico de Monterrey, Escuela de Ingeniería y Ciencias, México}%
\thanks{$^{7}$Viettel Cyberspace Center, Vietnam}%
\thanks{C. Authors: gilberto.ochoa@tec.mx, christian.daul@univ-lorraine.fr}%

\thanks{\textbf{*Preprint under review for EMBC'21 following IEEE guidelines}}}%

\begin{document}

\maketitle
\thispagestyle{empty}
\pagestyle{empty}

\begin{abstract}
Knowing the type (i.e., the biochemical composition) of kidney stones is crucial to prevent relapses with an appropriate treatment. During ureteroscopies, kidney stones are fragmented, extracted from the urinary tract, and their composition is determined using a morpho-constitutional analysis. This procedure is time consuming (the morpho-constitutional analysis results are only available after some days) and tedious (the fragment extraction lasts up to an hour). Identifying the kidney stone type only with the in-vivo endoscopic images would allow for the dusting of the fragments, while the morpho-constitutional analysis could be avoided. Only few contributions dealing with the in vivo identification of kidney stones were published. This paper discusses and compares five classification methods including deep convolutional neural networks (DCNN)-based approaches and traditional (non DCNN-based) ones. Even if the best method is a DCCN approach with  a precision and recall of 98\% and 97\% over four classes, this contribution shows that a XGBoost classifier exploiting well-chosen feature vectors can closely approach the performances of DCNN classifiers for a medical application with a limited number of annotated data.  

\end{abstract}

\indent \textbf{Keywords:} \textit {
Kidney stone recognition; endoscopy; deep learning; in vivo classification.}

\section{INTRODUCTION}

Kidney stones with a diameter of more than a few millimeters cannot usually leave the urinary tract, causing severe pain. During a standard ureteroscopy, kidney stones are visualized using digital ureteroscopes and broken into fragments using a laser. These fragments are extracted from the urinary tract and their biochemical constitution is analyzed in order to understand the causes (i.e. lithogenesis) leading to the formation of the kidney stones and to prevent relapses with an appropriate treatment (e.g., diet, drugs \cite{ friedlander2015}). The class of the extracted kidney stones  can be visually recognized by studying the textures, appearance and colours of the surfaces and sections of the fragments using a microscope. Complementary information about the crystalline composition can then be determined using infrared-spectrophotometry \cite{Khan2018}. However, in numerous hospitals, the result of such a morpho-constitutional analysis \cite{daudon2016} is usually available only one or two weeks after the ureteroscopy, and removing the kidney stone fragments is a tedious task that can last up to an hour.

Furthermore, only few and highly trained experts are able to recognize the type of a kidney stone by exclusively observing in vivo endoscopic images. A recent study \cite{estrade2020} has shown that the results of such visual recognition from endoscopic images by an expert is strongly correlated with those of the morpho-constitutional analyses. A visual in vivo type recognition in endoscopic images could save precious time since the fragments can be pulverized instead being extracted and the morpho-constitutional analysis can be avoided. However, most urologists are not trained to perform this kidney stone type recognition efficiently and such a task is also strongly operator dependent. In addition, it has been found that laser fragmentation can change the composition of kidney stones, which can bias the analyses \cite{keller2018}.

Despite the inherent advantages of an automated and objective kidney stone recognition method, only few studies have been published in this domain (see Table I for an overview of the literature). Both a classical approach (in \cite{serrat2017} a Random Forest classifier exploits histograms of RGB colours and local binary patterns -LBP- encoding rotation invariant textures) and a deep learning method \cite{torrell2018} (based on a Siamese Convolutional Neural Network, CNN) have been investigated, but they obtained rather moderate classification results (a mean accuracy of 63\% and 74\% was obtained over four and five classes for \cite{serrat2017} and \cite{torrell2018}, respectively). 
The authors in \cite{black2020} clearly improved the classification results on five kidney stone types using the ResNet-101 architecture (the leave-one-out cross-validation led to recall values from 71\% up to 94\% according to the class). The main limitation of these previous works lies in the fact that the methods were tested on ex-vivo images obtained in very controlled acquisition conditions and without endoscopes. In ureteroscopic in vivo data, the images are affected by blur, strong illumination changes between acquisitions, as well as by reflections, whereas the viewpoints are not easy to optimally adjust. However, these works have shown the feasibility of automating kidney stone classification.

\begin{table*}[t]
\centering
\caption{Overview of the data used in the literature of kidney stone classification. Six kidney stone types with different compositions were treated: Uric Acid (UA), Calcium Oxalate Monohydrate (COM), Calcium Oxalate Dihydrate (COD), Struvite (STR), Cystine (CYS) and Brushite (BRU). The classes and the acquisition conditions are given for each contribution.}
\begin{tabular}{lcccccccccc}
\hline
\multicolumn{1}{|l|}{\multirow{2}{*}{\begin{tabular}[c]{@{}l@{}}\\ References\end{tabular}}} & \multicolumn{6}{c|}{Kidney Stone Composition} & \multicolumn{2}{c|}{Image Type} & \multicolumn{2}{c|}{Acquisition Mode} \\ \cline{2-11} 
\multicolumn{1}{|l|}{} & \multicolumn{1}{l|}{UA} & \multicolumn{1}{l|}{COM} & \multicolumn{1}{l|}{COD} & \multicolumn{1}{l|}{STR} & \multicolumn{1}{l|}{CYS} & \multicolumn{1}{l|}{BRU} & \multicolumn{1}{l|}{Surface} & \multicolumn{1}{l|}{Section} & \multicolumn{1}{l|}{Ex Vivo} & \multicolumn{1}{l|}{In Vivo} \\ \hline
Serrat et al 2017 {[}5{]} & \ding{51}& \ding{51}& \ding{51}& \ding{51}& \ding{51}&  & \ding{51}& \ding{51}& \ding{51}&  \\ \hline
Torrell et al 2018 {[}6{]} & \ding{51}& \ding{51}& \ding{51}& \ding{51}& \ding{51}& \ding{51}& \ding{51}&  & \ding{51}&  \\ \hline
Black et al 2020 {[}7{]} & \ding{51}& \ding{51}&  & \ding{51}& \ding{51}& \ding{51}& \ding{51}& \ding{51}& \ding{51}&  \\ \hline
Martinez et al 2020 {[}8{]} & \ding{51}& \ding{51}& \ding{51}&  &  &  & \ding{51}& \ding{51}&  & \ding{51}\\ \hline
This contribution & \ding{51}& \ding{51}& \ding{51}&  &  & \ding{51}& \ding{51}& \ding{51}&  & \ding{51}\\ \hline
\end{tabular}
\label{pworks1}
\vspace*{-4mm}
\end{table*}

In a previous work  \cite{martinez2020}, it was shown that by choosing an appropriate colour space (HSI) and by extracting colour and texture features exploited by classical classifiers such as a Random Forest tree or a KNN ensemble model, the results can be significantly improved (an accuracy of about 85\% was obtained over 3 classes), even for patient data acquired with an endoscope in uncontrolled conditions.
The aim of this contribution is  to show how CNN-based solutions can further improve the classification of kidney stones acquired with ureteroscopes, even with a moderate number of images. 

The rest of this paper is structured as follows. Section II gives an overview of the acquired endoscopic image dataset and details the undertaken sampling and augmentation strategies to adapt the data to deep learning (DL) implementations. Section II presents also the model training approach. Section III compares the performances of the DL-strategies with those of conventional machine learning techniques. Finally, Section IV concludes the contribution and proposes perspectives.

\section{MATERIAL AND METHODS} \label{methods}

\subsection{Clinical Image Dataset}

\begin{figure}[b]
    \centering
    \includegraphics[width=210pt]{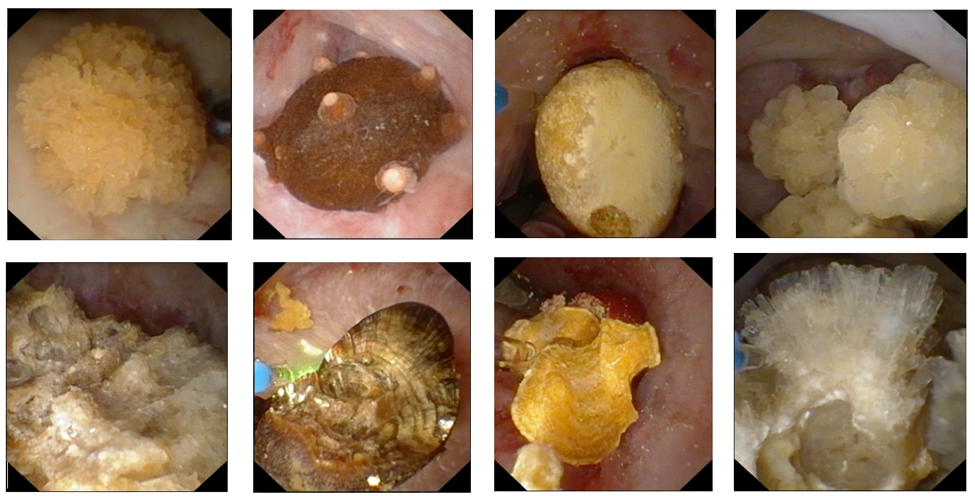}
    \caption{Examples of in vivo kidney stone images. From the left to the right: COM (whewellite), COD (weddelite), uric acid and brushite. Surface and section images are in the upper and lower line, respectively. }
    \label{fig:dataset}
\end{figure}

The dataset includes 177 kidney stone images which were acquired and annotated by an expert, Prof. Vincent Estrade (the data was collected following  the  ethical  principles  out-lined in the Helsinki Declaration of 1975, as revised in 2000, with the consent of the patients). The results of this visual classification were statistically confirmed by the concordance study in \cite{estrade2020}. The dataset consists of 90 fragment surface images and 87 fragment cross-section images of the four kidney stone types with the highest incidence (see the last row of Table \ref{pworks1}). These clinical images were captured using either the URF-V and URF-V2 endoscopes from  Olympus, or with a BOA endoscope from Richard Wolf. Images of this dataset are shown in Fig. \ref{fig:dataset}.

As in previous works [5-8], the classification is not performed on whole images, but using square patches localized on kidney stone surfaces or sections (surrounding tissues are not visible in the patches). In practice, kidney stone fragments have to be segmented prior to the classification. This process can be done either manually, or using automated methods, either in offline or online fashion.

\subsection{Patch extraction and data augmentation}

As confirmed by the results of previous works [5-8], image patches with a minimal size enable to capture enough texture and colour information for classification purposes. The use of image patches instead of the whole fragment surfaces and sections allows to increase the size of the training and test datasets. In order to avoid redundant information, the image areas including kidney stone fragments were scanned by square patches forming a regular grid whose neighbouring cells have a maximal overlap of twenty pixels. However, in previous works, the optimal size of these patches has not been studied. The patch size was a hyper-parameter which was adjusted during the training of the machine learning models presented in Section II.C.  The best size value was obtained after several ablation studies using four patch areas (64x64, 128x128, 256x256 and 512x512 pixels, respectively) and by monitoring the precision and loss curves for each patch size. The best trade-off in terms of accuracy and recall  was obtained with patches of 256x256 pixels. This patch size was used for the results given in Section III. As shown in Table II, 2680 surface patches and 2470 section patches were obtained in all with this procedure.

\begin{table}[b]
\centering
\begin{tabular}{|c|c|c|c|c|}
\hline
\multicolumn{2}{|c|}{Stone Type} & \multicolumn{2}{c|}{Acquired Images} &Number \\ \cline{1-4}
View & Class & Number & Presence (\%) & \begin{tabular}[c]{@{}c@{}}of patches\end{tabular} \\ \hline
\multirow{5}{*}{Surface} & COM & 30 & 31.9 & 870 \\ \cline{2-5} 
 & COD & 32 & 34.1 & 920 \\ \cline{2-5} 
 & Uric Acid & 18 & 19.1 & 470 \\ \cline{2-5} 
 & Brushite & 14 & 14.9 & 420 \\ \cline{2-5} 
 & \textbf{Total} & \textbf{94} & \textbf{100.0} & \textbf{2680} \\ \hline
\multirow{5}{*}{Section} & COM & 27 & 31.0 & 820 \\ \cline{2-5} 
 & COD & 28 & 32.2 & 780 \\ \cline{2-5} 
 & Uric Acid & 18 & 20.7 & 460 \\ \cline{2-5} 
 & Brushite & 14 & 16.1 & 410 \\ \cline{2-5} 
 & \textbf{Total} & \textbf{87} & \textbf{100.0} & \textbf{2470} \\ \hline
\end{tabular}
\label{patches_num}
\caption{Number of acquired images and of their (almost) non overlapping square patches }
\end{table}
As it can be observed in the last column of Table II, the resulting number of patches per class is imbalanced due to the changing fragment sizes and the number of images available per class. Two approaches were tested in order to balance the number of patches per class. In a ``down-sampling approach’’, the number of patches of each class is reduced to that of the brushite class with lowest sample number (420 and 410 patches for the surface and section images, respectively). 
In this ``down-sampling'' process, patches were randomly removed from the COM, COD and uric acid classes.  In contrast, in the ``up-sampling approach’’, the number of images of  the  brushite, COD and AU classes is increased to match that of the COM class with the highest sample number (870 and 820 patches for surface and section fragments, respectively). New patches, which are not located on the initial grid of patches, are randomly extracted from the images  to increase the sample number. Classification tests have shown that this ``up-sampling approach’’ led to slightly better results in terms of accuracy, which means that even if redundant information is present, the increase of the patch number favours the accuracy of the classification. With the ``up-sampling approach'', 750 patches are available per class.

Then, the number of patches of each class was still increased since the performance of deep learning (DL) approaches relates to the amount of available training data. The data was augmented by applying different combinations of geometrical transformations to the original patches: patch flipping, affine transformations, and perspective distortions. The number of patches increased from 5,400 to 43,200 using this data augmentation (10\% of the original patches were kept for test purposes). The patches were also ``whitened’’ using the mean $m_i$ and standard deviation $\sigma_i$ of the colour values $I_i$ in each channel ($I_i^w = (I_i – m_i)/\sigma_i$, with i = R, G, B).  This dataset was split in three parts for the training, validation and test stages.
\subsection{Feature extraction and classification}
The aim of this paper is notably to compare the deep-leaning methods to the best ``classical’’ classification methods (i.e., non-DL based approaches). In [8], the feature vector (based on HSI colour energies and rotation invariant LBP histograms) was identified as leading to the highest separability of the kidney stones. Among the classical methods exploiting these features, Random Forest trees and XGBoost classifiers obtained the highest precision and recall (P and R, respectively in Table III) for in vivo kidney-stone images. For these two classical machine learning methods, the results given in Section III were obtained by a hyper-parameter tuning using a 10 fold cross-validation (CV) and by averaging the results over five runs (for the non DL models we used on-leave-out CV due to the low number of samples).

 DL architectures of various levels of complexity (AlexNet, VGG16 and Inception v3) were adapted for this contribution. The performance of the feature extractor backbones of these models is optimal only when a large number of training images is available. The proposed method leverages the benefits of transfer learning by exploiting CNN backbones pre-trained with ImageNet. The fully connected (FC) layers of the original backbones are replaced by a custom FC layer of 25 channels, followed by a Batch Normalization, a ReLU activation function, another FC layer of 256 channels and a softmax layer with 4 class outputs. The two FC layers were randomly initialized and connected to a softmax layer for predicting the patch class. During the training process of the three DL models using the kidney stone patches, the weights in the convolution layers were fixed and only the weights in the FC layers were updated. 

For all the reported experiments, we made use of Pytorch 1.7.0 and CUDA 10.1.  The learning rates for each model were obtained using the Pytorch Lightning 1.0.2 optimizer,  yielding the following learning rate values: 0.0001 (AlexNet), 0.00005 (VGG16) and 0.0006 (Inception V3). We used the ADAM optimizer, a batch size of 64 and early stopping for all the experiments, whose results are discussed in the next section.

\begin{table}[]
\fontsize{8}{9}\selectfont
\centering
\begin{tabular}{|c|c|c|c|c|c|c|}
\hline
\multirow{2}{*}{Classifier} & \multicolumn{2}{c|}{Surface} & \multicolumn{2}{c|}{Section} & \multicolumn{2}{c|}{Mixed} \\ \cline{2-7} 
 & P & R & P & R & P & R \\ \hline
R. Forest & 0.87 & 0.82 & 0.82 & 0.82 & \textbf{0.91} & \textbf{0.91} \\ \hline
XGboost & 0.93 & 0.93 & 0.89 & 0.89 & \textbf{0.96} & \textbf{0.96} \\ \hline
AlexNet & 0.93 & 0.95 & 0.83 & 0.82 & 0.92 & 0.92 \\ \hline
VGG19 & 0.95 & 0.96 & 0.91 & 0.92 & 0.94 & 0.92 \\ \hline
Inception & \textbf{0.98} & \textbf{0.97} & \textbf{0.94} & \textbf{0.96} & \textbf{0.97} & \textbf{0.98} \\ \hline
\end{tabular}
\caption{Weighted average metrics comparison for section and surface patches, as well as mixed patches.}
\label{dl_results1}
\end{table}

\begin{table}[!t]
\fontsize{7.5}{9}\selectfont
\centering
\begin{tabular}{|c|c|c|c|c|c|c|c|c|}
\hline
\multirow{2}{*}{Classifier} & \multicolumn{2}{c|}{COM} & \multicolumn{2}{c|}{COD} & \multicolumn{2}{c|}{UA} & \multicolumn{2}{c|}{BRU} \\ \cline{2-9} 
 & P & R & P & R & P & R & P & R \\ \hline
R. Forest & 0.84 & 0.86 & 0.90 & 0.95 & 0.88 & 0.67 & 0.90 & 0.92 \\ \hline
XGBoost & 0.92 & 0.96 & 0.91 & 0.91 & 0.97 & 0.96 & 0.96 & 0.94 \\ \hline
AlexNet & 0.93 & 0.98 & 0.95 & 0.85 & 0.88 & 0.92 & 0.93 & 0.92 \\ \hline
VGG19 & 0.97 & 0.97 & 0.92 & 0.93 & 0.93 & 0.83 & 0.94 & 0.92 \\ \hline
Inception & \textbf{0.98} & \textbf{0.97} & \textbf{0.93} & \textbf{0.96} & \textbf{0.95} & \textbf{0.90} & \textbf{0.96} & \textbf{0.95} \\ \hline
\end{tabular}
\caption{Precision (P) and recall (R) values obtained over the four classes with all classifiers.  }
\label{dl_results2}
\vspace*{-7mm}
\end{table}
\section{RESULTS AND DISCUSSION} \label{results}
Various experiments were carried out for evaluating the machine learning methods described in section II.C using the patch data introduced in II.B. In particular, the ability of these models to predict the kidney stone class either based on surface and section patches taken separately, or by using simultaneously the two patch types, as is it classically done in the morpho-constitutional analysis procedure \cite{daudon2016}. To do so, all models were trained three times, namely i) solely with the section patches, ii) only with the surface patches, and iii) by mixing the two patches types. Each section or surface patch was classified two times, i.e. by its dedicated model and by the model for mixed data. The well-known precision (P, see Tables III and IV) and recall (R) metrics are determined for each kidney stone class to assess the clinical applicability of the studied fragment classification methods. 

The results obtained for the Random Forest classifier (see the first rows of Tables III and IV) led to very similar performances as those reported in \cite{martinez2020}, showing that the class balancing compensates the increase of the number of classes. Moreover, still with respect to [8], an additional classifier was tested: XGBoost yielded significantly better results than the Random Forest classifier. These results are even comparable to those obtained with the DL-AlexNet model.

The results of the chosen deep learning (DL) models are summarized in the last three rows of Tables III and IV. It is noticeable that only the Inception v3 model outperforms the best traditional machine learning method based on the XGBoost classifier. As seen in the last rows of Table IV, this model exhibits the highest mean precision (P) and recall (R) for all classes and for all patch types (section patches, surface patches and mixed patch types).

 It can be observed in Table III that for the classical models, the simultaneous use of surface and section patches lead to the best results, whereas the three DL-based methods exhibit globally the best overall results when exploiting only surface images (confirming the results of the concordance study in \cite{estrade2020}). However, among the DL approaches, only Inception v3 has globally a better precision (P = 0.97) and recall (R = 0.98) than the XGBoost classifier (P = R = 0.96). 
 
 \begin{figure}[t]
    \centering
    \includegraphics[width=180pt]{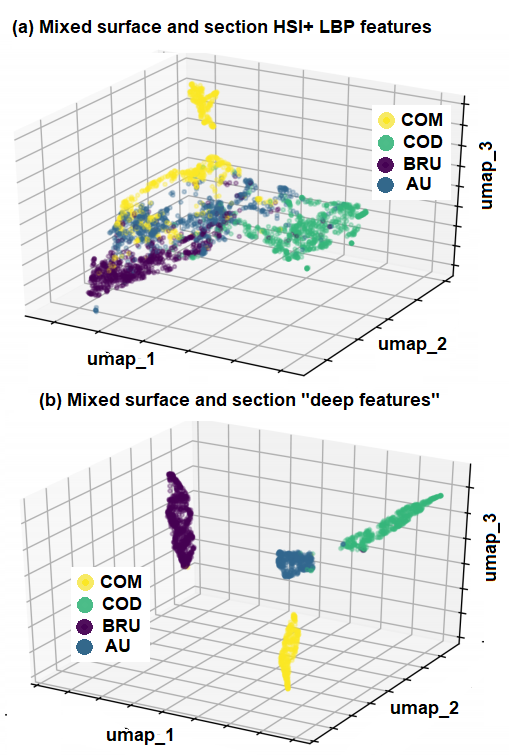}
    \caption{Feature visualization using the UMAP dimensionality reduction technique for (a) the HSI and LBP features and (b) the ``deep features".}
    \label{fig:features}
    \vspace*{-5mm}
\end{figure}

Figure 2(a) provides an UMAP visualisation \cite{mcinnes2020} which illustrates the class separability achieved using only the three most discriminant dimensions (umap1 to umap3) obtained after the dimensionality reduction of the HSI-LBP feature space.  In Fig. 2(b) it is noticeable  that the same UMAP dimensionality reduction applied on the “deep features”  produce tighter clusters and larger inter-class distances than in Fig.~2(a). This suggests that the superior performance of the DL models relates to the richness of the information extracted from the patches using efficient feature extraction backbones associated with transfer learning. 

Inception v3 and XGBoost improve the precision and recall metrics by about 30\% and 10\% in comparison to the traditional approaches in [5] and [8], respectively. These two models outperform also the DL model tested in [7] on ex-vivo data, which exhibited a high average recall of 97\%, but a lower average precision of 80\%.

\section{CONCLUSION} \label{conclusions}

In this work, it was shown that it is possible to effectively train DL models for predicting kidney stone types using a rather small dataset acquired during ureteroscopies. This study confirms that AI-methods can be incorporated in the urologists’ workflow for identifying the causes of the kidney stone formation \cite{jah2020}. Thus, the tedious fragment extraction phase can be avoided since pulverizing kidney stones becomes possible without losing the information necessary for diagnosis purposes. However, this work focused only on four classes of kidney stones. The proposed method should be improved to identify classes with a lower incidence and/or mixed composition.  Additionally, all previous works including ours make use of ``still’’ images. Classifying kidney stones using video data could further facilitate the urologist's work and represents a fertile area of research, but poses very complex challenges, as the regions of interest might be blurred due to the endoscope motion, affected by reflections, or occluded by surgical instruments, blood or debris.  

\section*{Acknowledgments}
The authors wish to thank the AI Hub and the CIIOT at ITESM for their support for carrying the experiments reported in this paper in their NVIDIA's DGX computer.

We also thank the Verano de la Investigación Cientifica Delfin program for assisting Andres Varela with a mobility grant. and CONACYT for the master scholarhip for Oscar Hinojosa at Universidad Autonoma de Guadalajara (Mexico).

\addtolength{\textheight}{-12cm}   


\end{document}